\documentclass[twoside,11pt]{article}

\usepackage[accepted]{melba}

\usepackage{mwe} 

\usepackage{amsmath,amsfonts}

\melbaid{2024:019}  
\doi{https://doi.org/10.59275/j.melba.2024-9c68}
\melbaauthors{Newlin}  
\volume{2}
\firstpageno{1083}  
\melbayear{2024}  
\datesubmitted{06/2024}  
\datepublished{08/2024}  

\melbaspecialissueeditors{Marleen de Bruijne, Tal Arbel, Ismail Ben Ayed, Hervé Lombaert}

\ShortHeadings{QuantConn challenge findings}{Newlin}

\title{MICCAI-CDMRI 2023 QuantConn Challenge Findings on Achieving Robust Quantitative Connectivity through Harmonized Preprocessing of Diffusion MRI}

\author{\firstname Nancy R. \surname Newlin \email nancy.r.newlin@vanderbilt.edu  
	\addr CS, Vanderbilt University, Nashville, TN, USA
}
\author{{Nancy R. Newlin$^*$$^1$,
    Kurt Schilling$^2$,
    Serge Koudoro$^3$,
    Bramsh Qamar Chandio$^4$,
    Praitayini Kanakaraj$^1$,  
    Daniel Moyer$^1$,  
    Claire E. Kelly$^{5,6,7}$,
    Sila Genc$^{5,8}$, 
    Jian Chen$^{5}$,
    Joseph Yuan-Mou Yang$^{5,8,9,10}$,  
    Ye Wu$^{11}$, 
    Yifei He$^{11}$, 
    Jiawei Zhang$^{12}$, 
    Qingrun Zeng$^{13}$, 
    Fan Zhang$^{13}$,
    Nagesh Adluru$^{14}$,   
    Vishwesh Nath$^{15}$,  
    Sudhir Pathak$^{16}$,  
    Walter Schneider$^{16}$,  
    Anurag Gade$^{17}$,    
    Yogesh Rathi$^{18}$, 
    Tom Hendriks$^{19}$,   
    Anna Vilanova$^{19}$,  
    Maxime Chamberland$^{19}$,  
    Tomasz Pieciak$^{20,21}$,  
    Dominika Ciupek$^{21}$,  
    Antonio Tristán Vega$^{20}$, 
    Santiago Aja-Fernández$^{20}$,  
    Maciej Malawski$^{21}$,  
    Gani Ouedraogo$^{22}$,  
    Julia Machnio$^{21}$,  
    Christian Ewert$^{23}$,
    Paul M. Thompson$^4$,  
    Neda Jahanshad$^4$,  
    Eleftherios Garyfallidis$^3$,  
    Bennett A. Landman$^{1,2,24,25}$
    } \\
    \scriptsize{
        $^{1}$Department of Computer Science, Vanderbilt University, Nashville, TN
$^{2}$Department of Radiology and Radiological Sciences, Vanderbilt University Medical Center
$^{3}$Indiana University Bloomington, Bloomington, IN
$^{4}$Mark and Mary Stevens Neuroimaging and Informatics Institute, Keck School of Medicine of USC, Los Angeles, CA
$^{5}$Developmental Imaging, Murdoch Children’s Research Institute, Melbourne, Australia
$^{6}$Victorian Infant Brain Studies (VIBeS), Murdoch Children’s Research Institute, Melbourne, Australia
$^{7}$Turner Institute for Brain and Mental Health, School of Psychological Sciences, Monash University, Melbourne, Australia
$^{8}$Neuroscience Advanced Clinical Imaging Service (NACIS), Department of Neurosurgery, Royal Children’s Hospital, Melbourne, Australia
$^{9}$Neuroscience Research, Murdoch Children’s Research Institute, Melbourne, Australia
$^{10}$Department of Pediatrics, University of Melbourne, Melbourne, Australia
$^{11}$School of Computer Science and Technology, Nanjing University of Science and Technology, Nanjing, China
$^{12}$College of Information Engineering, Zhejiang University of Technology, Hangzhou, China
$^{13}$School of Information and Communication Engineering, University of Electronic Science and Technology of China, Chengdu, China
$^{14}$Waisman Center, Department of Radiology, University of Wisconsin, Madison
$^{15}$NVIDIA, Nashville, TN, USA
$^{16}$Learning Research and Development Center, University of Pittsburgh
$^{17}$Brigham and Women’s Hospital, Boston
$^{18}$Brigham and Women's Hospital, Harvard Medical School, Boston
$^{19}$Department of Computer Science and Mathematics, Eindhoven University of Technology, Netherlands
$^{20}$LPI, ETSI Telecomunicación, Universidad de Valladolid, Castilla y León, Spain
$^{21}$Sano Centre for Computational Medicine, 30-054 Kraków, Poland
$^{22}$Aix-Marseille Université, Marseille, France
$^{23}$ AI in Medical Imaging, German Center for Neurodegenerative Diseases (DZNE)
$^{24}$Vanderbilt University Institute of Imaging Science, Vanderbilt University, Nashville, TN, USA
$^{25}$Department of Electrical and Computer Engineering, Vanderbilt University, Nashville, TN, USA
}	
\\\\Corresponding author: \url{nancy.r.newlin@vanderbilt.edu}
}    
\begin{document}

\maketitle

\begin{abstract}
	White matter alterations are increasingly implicated in neurological diseases and their progression. International-scale studies use diffusion-weighted magnetic resonance imaging (DW-MRI) to qualitatively identify changes in white matter microstructure and connectivity. Yet, quantitative analysis of DW-MRI data is hindered by inconsistencies stemming from varying acquisition protocols. Specifically, there is a pressing need to harmonize the preprocessing of DW-MRI datasets to ensure the derivation of robust quantitative diffusion metrics across acquisitions. In the MICCAI-CDMRI 2023 QuantConn challenge, participants were provided raw data from the same individuals collected on the same scanner but with two different acquisitions and tasked with preprocessing the DW-MRI to minimize acquisition differences while retaining biological variation. Harmonized submissions are evaluated on the reproducibility and comparability of cross-acquisition bundle-wise microstructure measures, bundle shape features, and connectomics. The key innovations of the QuantConn challenge are that (1) we assess bundles and tractography in the context of harmonization for the first time, (2) we assess connectomics in the context of harmonization for the first time, and (3) we have 10x additional subjects over prior harmonization challenge, MUSHAC and 100x over SuperMUDI.  We find that bundle surface area, fractional anisotropy, connectome assortativity, betweenness centrality, edge count, modularity, nodal strength, and participation coefficient measures are most biased by acquisition and that machine learning voxel-wise correction, RISH mapping, and NeSH methods effectively reduce these biases. In addition, microstructure measures AD, MD,  RD, bundle length, connectome density, efficiency, and path length are least biased by these acquisition differences. A machine learning approach that learned voxel-wise cross-acquisition relationships was the most effective at harmonizing connectomic, microstructure, and macrostructure features, but requires the same subject be scanned at each site co-registered. NeSH, a spatial and angular resampling method, was also effective and has generalizable framework not reliant co-registration.
	Our code is available at~\url{https://github.com/nancynewlin-masi/QuantConn/}.
\end{abstract}

\begin{keywords}
	Diffusion MRI, harmonization, tractometry, tractography, connectomics, image processing
\end{keywords}

\section{Introduction}
Diffusion-weighted magnetic resonance imaging (DW-MRI) of the brain enables in-vivo characterization of white matter microstructure and supports structural brain connectivity mapping \citep{Pierpaoli1996}. DW-MRI acquisition involves varying magnetic field strength with pulsed gradients to sensitize to the movement of water molecules, following the Stejskal and Tanner method \citep{Stejskal1965}. There are a growing number of multi-site DW-MRI studies that encompass varying scanner manufacturers and acquisition protocols. An imaging “site” refers to the acquisition protocol parameters and scanner specifications used to collect an image. Initiatives such as the Alzheimer’s Disease Neuroimaging Initiative (ADNI) \citep{Jack2008}, the National Alzheimer’s Coordinating Center (NACC) \citep{NACC}, the Open Access Series of Imaging Studies (OASIS3) \citep{LaMontagne2019}, and the Baltimore Longitudinal Study of Aging (BLSA) \citep{Ferrucci2008} incorporate data from diverse scanner vendors and protocols.

Diffusion imaging inherits site-effects of conventional MRI caused by magnetic field inhomogeneities, field strength, voxel size, and vendor differences. Additionally, it has unique challenges related to diffusion sensitization and processing, including the number of directions the gradient field is applied, the timing of applied gradients, and the DW-MRI reconstruction algorithm \citep{Nencka2018, Ni2006}. Perturbations to these acquisition decisions within and across datasets can introduce significant confounding site-dependent differences in DW-MRI and subsequent connectomic and bundle analyses. Vollmar et al. demonstrated confounding site differences in the analysis of whole brain, region of interest, and tract-defined microstructure in a cohort of traveling subjects scanned with various protocols and scanner vendors \citep{Vollmar2010}. Similar findings emerged from studies involving multiple vendors, models, and protocols, leading to substantial sources of variation  \citep{Karayumak2019}. This variability extends to derived metrics, such as tractography bundles \citep{Schilling2021} and complex network measures \citep{Newlin2023, Onicas2022}. Schilling et al. indicated that fiber bundle shape and microstructure analysis were influenced by scanner manufacturer, acquisition protocol, diffusion sampling scheme, diffusion sensitization, and overall bundle processing workflow \citep{Schilling2021}. Joint datasets from retrospective studies by Newlin et al. and Onicas et al. revealed significant differences in complex network measures, such as modularity, global efficiency, clustering coefficient, density, characteristic path length, small-worldness, and average betweenness centrality, attributed to variations in protocol and scanner vendor \citep{Newlin2023, Onicas2022}.

Consequently, there is a clear  imperative to address these site-effects in connectivity and structure analyses through a process commonly known as "harmonization". Diffusion image harmonization refers to methods,  using preprocessing, machine learning, resampling, etc., that reduce bias associ ated with data collection and storage while preserving biological variation \citep{Pinto2020}.

The Quantitative Connectivity through Harmonized Preprocessing of Diffusion MRI (QuantConn) challenge is intended to detail and evaluate current image acquisition harmonization techniques. Our evaluation is focused on reproducibility of downstream tasks (tractography, connectomics, tractometry) and their features (bundle macrostructure and microstructure, complex network measures of the connectome). 

\section{Related Works}
   \begin{figure}[h]
		\centering
		\includegraphics[width=1.0\linewidth]{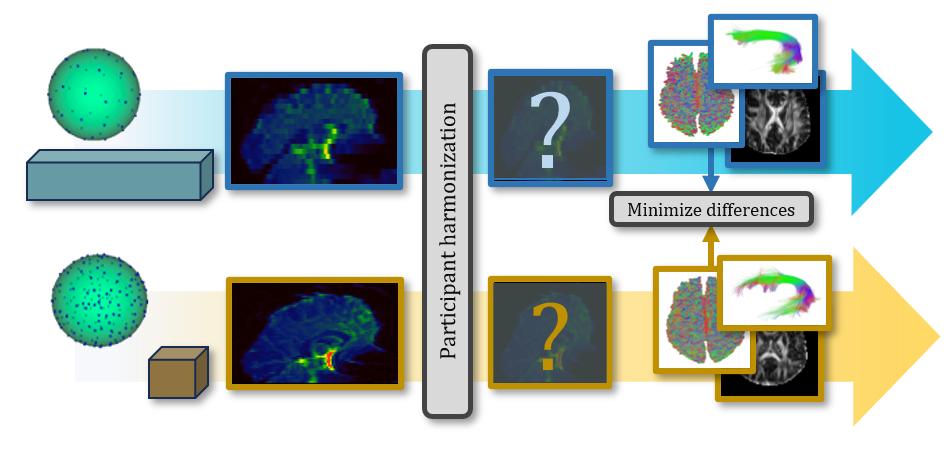}
		\caption{We released 206 scans across two acquisitions, “A” (blue) and “B” (orange). Acquisition A was acquired with anisotropic resolution and 27 gradient directions. Acquisition B was acquired with isotropic resolution and 94 gradient directions. Participants altered the DW-MRI with the harmonization of their choosing. We then feed this harmonized data through a standard processing pipeline of tensor fitting, orientation distribution function (ODF) estimation, tractography, and tractometry to arrive at diffusion metrics of bundles microstructure and macrostructure and connectomics. Harmonization efficacy is determined by its capacity to minimize differences in these downstream diffusion measures.}
    \end{figure}
    QuantConn builds off the successful SuperMUDI  \citep{Pizzolato2020} and MUSHAC  \citep{Ning2019} challenges. SuperMUDI challenged the community with overcoming acquisition differences in voxel anisotropy with superresolution. The study consisted of N=5 volunteers with high in-plane resolution and lower axial resolution. In the MUSHAC challenge, participants were given datasets (N=15 volunteers) that were acquired on two different scanners and two protocols on each and tasked with minimizing cross-scanner and cross-protocol differences in voxel-wise indices. We provide 103 pairs of datasets of the same subjects scanned with different acquisition protocols - and challenge participants to minimize differences in the data in order to minimize differences in microstructure, tractography bundle analysis, and connectomics measures (Figure 1). The QuantConn challenge introduces several important advancements. Firstly, it is the first to evaluate bundles, tractography, and connectomics within the context of harmonization. It features a dataset that includes ten times more subjects than MUSHAC and a hundred times more than SuperMUDI. The data that form the basis of this challenge represent a difficult clinical scenario for harmonization and are part of a much larger twins study, which could provide a rich context for continuing validation and extension of this challenge's findings  \citep{Queensland}.

\section{Methods}
	Data includes 103 patients, scanned twice with acquisition protocol “A” and “B”. Challenge participants created or applied harmonization method of their choice to bridge the gap between acquisitions (Table 1). Submissions' harmonized data was processed using a standard diffusion pipeline of tensor fitting, tissue segmentation, tractography, connectomics, and tractometry (Figure 2). Then, we evaluated the cross-acquisition agreement of diffusion features resulting from this pipeline.	\\

\begin{table}

\caption{Summary of submissions (Sub) and their 3 scores. Scores are computed by taking the average intra-class correlation coefficient (ICC) (rater=acquisition, measurements=25 test subjects). For connectomics (Conn), we averaged ICC for all 10 complex network measures. For microstructure (Micro), we averaged ICC for all 6 bundles and 4 measures. For macro (Macro), we averaged ICC for all 6 bundles and 6 measures. See supplementary information for in-depth analysis of the stability of rankings for each metric individually.}
\resizebox{\textwidth}{!}{
\begin{tabular}{c|p{21 em}ccc}
Sub & Method                                                                 & Conn Score & Micro Score & Macro Score 
\\
\hline
Ref                         & None                                             & 0.79$\pm$0.14 & 0.24$\pm$0.23 & 0.49$\pm$0.28 \\
1                           & PreQual, registration, voxel-wise map with MLP   & 0.98$\pm$0.05 & 0.79$\pm$0.25 & 0.70$\pm$0.29 \\
2                           & Normalization, RISH feature map                  & 0.84$\pm$0.16 & 0.49$\pm$0.21 & 0.62$\pm$0.28 \\
3                           & PreQual, Spatial and angular resampling                   & 0.79$\pm$0.16 & 0.69$\pm$0.11 & 0.45$\pm$0.32 \\
4                           & PreQual, registration, DeepHarmony               & 0.86$\pm$0.08 & 0.36$\pm$0.26 & 0.37$\pm$0.26 \\
5                           & PreQual, bias field correction, RISH feature map & 0.81$\pm$0.10 & 0.30$\pm$0.12 & 0.48$\pm$0.28 \\
6                           & PreQual                                          & 0.75$\pm$0.18 & 0.29$\pm$0.19 & 0.45$\pm$0.26 \\
7                           & PreQual, registration, MLP with dynamic input    & 0.75$\pm$0.14 & 0.24$\pm$0.20 & 0.44$\pm$0.28 \\
8                           & PreQual, map SH coefs from acquisition A to B    & 0.24$\pm$0.24 & 0.46$\pm$0.23 & 0.07$\pm$0.11 \\
9                           & PreQual, map SH coefs from acquisition B to A    & 0.18$\pm$0.24 & 0.40$\pm$0.20 & 0.06$\pm$0.10      
\end{tabular}
}
\end{table}

	\subsection{Data}
		The data that form the basis of this challenge represent a difficult clinical scenario for harmonization and are a subset of the Queensland Twin Imaging study  \citep{Queensland}. The DW dataset consists of 25 testing and 78 training subjects scanned twice with two different acquisition protocols for a total of 206 scanning sessions. Each subject has an anatomical T1-weighted image. The data subset is comprised of 45\% females, ages 25.3 $\pm$ 1.8 years. No subjects reported a history of significant head injury, neurological or psychiatric illness, or substance abuse or dependence, and no subjects had a first-degree relative with a psychiatric disorder. All subjects were right-handed as determined using 12 items from Annett’s Handedness Questionnaire  \citep{ANNETT1970}. Scanning was performed at the QIMR Berghofer Medical Research Institute on a 4 Tesla Siemens Bruker Medspec scanner.
        \subsubsection{Anatomical Imaging}
        T1-weighted images were acquired with an inversion recovery rapid gradient-echo sequence (inversion/repetition/echo times, 700/1500/3.35 ms; flip angle, 8 degrees; slice thickness, 0.9 mm; 256 $x$ 256 acquisition matrix). 

        \subsubsection{Acquisition A}
            DW images were acquired using single-shot echo-planar imaging with a twice-refocused spin echo sequence. Imaging parameters were repetition/echo times of 6090/91.7 ms, field of view of 23 cm, and 128 × 128 acquisition matrix. Each 3D volume consisted of 21 axial slices 5 mm thick with a 0.5 mm gap and 1.8 × 1.8 $mm^2$ in-plane resolution, total time = 3 minutes. Thirty images were acquired per subject: three with no diffusion sensitization (b=0 $s/mm^2$) and 27 DW images (b = 1146 $s/mm^2$ ) with gradient directions uniformly distributed on the hemisphere.
        \subsubsection{Acquisition B}
            DW images were acquired using single-shot echo planar imaging (EPI) with a twice-refocused spin echo sequence. Imaging parameters were: 23s cm FOV, TR/TE 6090/91.7 ms, with a 128 × 128 acquisition matrix. Each 3D volume consisted of 55 2-mm thick axial slices with no gap and a 1.79 × 1.79 $mm^2$ in-plane resolution with total acquisition time = 14.2 minutes. 105 images were acquired per subject: 11 with no diffusion sensitization (b=0 $s/mm^2$) and 94 DWI (b = 1159 $s/mm^2$) with gradient directions uniformly distributed on the hemisphere.
    \subsection{Diffusion Processing}
        \begin{figure}
    \centering
    \includegraphics[width=1.0\linewidth]{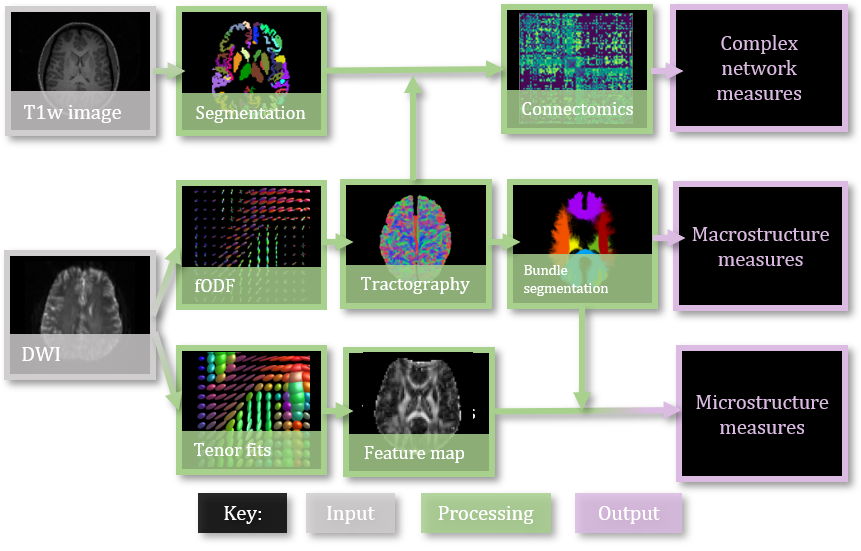}
    \caption{Using the harmonized data provided by participants, the full testing pipeline is as follows: tensor fitting, fODF estimation, whole brain tractography, bundle segmentation and tractometry, then connectomics, and finally complex network analysis.
These processes result in three groups of analysis: complex network measures, bundle microstructure, and bundle macrostructure, which we evaluate the submissions on.
}
    \end{figure}
        Tensor models are fit to the log-signal by first least squares (WLS) weighted by empirical signal intensities, and then iterative WLS using the previous step’s signal prediction  \citep{Basser1994}. Then, we generate whole-brain maps of tensor parameters: mean diffusivity (MD), fractional anisotropy (FA), axial diffusivity (AD), radial diffusivity (RD). The tensor has three eigenvalues: one principal, equivalent to AD, and two non-principal, the mean of which is RD. MD is the average magnitude of observed diffusion  \citep{Westin2002}. FA is a scaled variability measure of diffusivity  \citep{Westin2002}.
        
        We ran Freesurfer on all T1-weighted images to get anatomical parcellations. This provides a 5-tissue type image, grey-matter white matter boundary region segmentation, and an 84-node grey-matter parcellation  \citep{Fischl2012}.The grey-matter parcellation, defined by the Desikan-Killany grey-matter atlas, is used to map tractography to anatomical regions for connectomics analysis  \citep{Klein2012}. 
        
        From the image signal, we can estimate white matter fiber orientation distribution functions using constrained spherical deconvolution  \citep{Tournier2013,Tournier2007}. fODF models reflect the probability that a real brain fiber lies along a given direction  \citep{Tournier2007}. These models guide tractography. Seeding from the grey-matter white matter boundary, we perform second order integration over fODFs until 10 million streamlines are sampled  \citep{Newlin2023, mrtrix}. 
        
        We segment the tractogram into 30 bundles with RecoBundlesX  \citep{Garyfallidis2018}, but only keep 6 major tracks for analysis: bilateral Arcuate Fasciculus (AF), bilateral Optic Radiations (OR), and major/minor corpus colosum forceps. These bundle microstructure profiles of this bundles are created by projecting DTI maps onto the bundles using BUAN approach \cite{Chandio2020}, bundle profiles are averaged and result in average FA, MD, RD, and AD for each bundle. In addition, we characterize bundles by their macrostructure and shape. Here, we consider bundle length, volume, curl, span, diameter, and surface area  \citep{scilpy}.
        
        We combine the tractogram with the 84 node grey-matter parcellation from Freesurfer, Deskian-Killany  \citep{Fischl2012}. For each region $i$, we compute the number of streamlines and average length of streamline connecting region $i$ to every other region. We store this value in an adjacency matrix called a connectome  \citep{mrtrix}. 
        
        The connectome is comprised of nodes, brain regions defined by the Desikan-Killany atlas, and edges, a scalar value reflecting the strength of the connection. The brain connectivity toolbox (version-2019-03-03)  \citep{bct} specializes in complex network measures that summarize the connectivity behavior of a connectome. We consider 12 complex network measures as connectomics measures. The importance of brain regions in the network is measured with node strength and betweenness centrality. Connectome density is the number of connections found versus the total possible connections. Functional integration is measured with characteristic path length. Characteristic path length, edge count, and global and local efficiency measure the ability to exchange information. Modularity and participation coefficient describe the community structure of the brain network. Assortativity measures the resiliency of a network to connection drop out. Together, these complex network measures characterize brain structural connectivity. 

    \subsection{Harmonization Methods}
        The QuantConn challenge received a total of 9 unique submissions (Table 1). The following section details the preprocessing and harmonization methodology used in each submission.  
        
            \subsubsection{Baseline}
            DW-MRI was not preprocessed.
            No harmonization method was applied.
            \subsubsection {Submission 1: The Harmonizers 1}
            All DW-MRI were processed using the PreQual pipeline to remove eddy current, motion, and echo-planar imaging (EPI) distortions  \citep{Cai2021}. Additionally, average b=0 images of the subject’s diffusion data of acquisition A and B were mapped to the subject’s T1 image with boundary-based rigid registration  \citep{Greve2009} after running FreeSurfer \citep{Fischl2012} on the T1 images. The obtained mappings were used to map the subject’s acquisition A data to the respective acquisition B data, where the mapped data was regridded to the acquisition B target resolution.
            For the harmonization step, an MLP was trained to summarize the acquisition A and acquisition B in a voxel-wise manner. The subject’s data from both sites were concatenated along the channel dimension per voxel for input. The output format was chosen based on acquisition B to leverage its inclusion of more b-vectors, potentially enhancing tractography. The neural network was an MLP with one hidden layer containing 4096 units, batch normalization, and a ReLU activation. The training set included 75 subjects, whereas the validation set included 3 subjects. The network was trained for 40 epochs with a mean squared reconstruction loss using the learning rate 0.01.

            \subsubsection{Submission 2: NIMG}
            The DW-MRI scans were processed using two software tools, MRtrix3 \cite{mrtrix} and FSL \cite{fsl}, to eliminate magnetic field inhomogeneities, signal inhomogeneities, and volume drift across the diffusion gradients. The harmonization of the DW-MRI data involved three major steps. Firstly, a global DW-MRI intensity normalization was performed on a group of subjects using MRtrix3, with the median b=0 white matter value serving as the reference. Secondly, a joint eddy current-induced distortion correction was applied to the concatenated DWI data over the two sites per subject. Thirdly, an individual DW-MRI intensity normalization was performed using MRtrix3 for each subject based on the b=0 signal. Finally, an algorithm called the DW-MRI harmonization algorithm was employed. This algorithm leverages rotation invariant spherical harmonics (RISH) features to construct a scale map for each pair of reference and target sites. The scale map was then applied to the target site to remove inter-site variability \citep{Karayumak2019,Mirzaalian2018}. The code is open source  \citep{linearrish}.

            \subsubsection{Submission 3: NeSH}
            All DW-MRI were processed using the PreQual pipeline to remove eddy current, motion, and EPI distortions  \citep{Cai2021}. Additionally, DW-MRI was noise corrected and de-gibbs ringing with MRTrix3  \citep{mrtrix}.
            DW-MRI were harmonized by resampling spatial and angular resolutions to be the same between acquisitions. Spatial resolution was set to be 1.79688 isotropic. The sampled directions were a combination of the directions used in site A and site B. The 27 original b-vectors for A were used, alongside 33 b-vectors from B that were least similar to the b-vectors from A. The resampling process used NeSH to create continuous implicit neural representations of the DW-MRI that can be sampled in any spatial and angular resolution \citep{HendricksCDMRI2023}. Optimal training parameters were batch size=1000, epochs=7, learning rate=$1\times10^{-6}$, layers=8, lambda=$1\times10^{-6}$, sigma=4.
            
            \subsubsection{Submission 4: DiffusionMaRInes}
            All DW-MRI were processed using the PreQual pipeline to remove eddy current, motion, and echo-planar imaging (EPI) distortions \citep{Cai2021}. All non-diffusion-weighted MR data were averaged into a single volume.
            This submission harmonized the angular resolution of acquisition B to match acquisition A and the spatial resolution of acquisition B to match acquisition A. The datasets from acquisition A to acquisition B were spatially interpolated using a trilinear interpolation. All harmonized DW-MRI had 27 gradient directions and 1.8 mm isotropic resolution. The resampling process used the DeepHarmony model  \citep{Dewey2019} with (epochs=20, batch size=8, features=16, Adam beta1=0.9, Adam beta2=0.999, patch size=64x64). The training procedure involved 10 randomly selected subjects from the training set using 6-fold cross-validation. Images from acquisition A were linearly registered to their acquisition B counterpart. The registration transform was computed on respective FA maps with 6 degrees of freedom and optimized using mutual information \citep{fsl}.  
            
            \subsubsection{Submission 5: Neuroscience Advanced Clinical Imaging Service (NACIS)}
            All DW-MRI were processed using the PreQual pipeline to remove eddy current, motion, echo-planar imaging (EPI) distortions  \citep{Cai2021}, and B1-bias field correction using ANTs N4.
            DW-MRI signal was harmonized by mapping rotationally invariant features of the spherical harmonic representation to a common domain (acquisition B)  \citep{Karayumak2019,Mirzaalian2018}. The code is open source  \citep{linearrish}. This method consists of creating a template mapping between the rotationally invariant features with N=10 matched subject pairs from both sites. 
            \subsubsection{Submission 6: PreQual}
            All DW-MRI were processed using the PreQual pipeline to remove eddy current, motion, and EPI distortions  \citep{Cai2021}. 
            \subsubsection{Submission 7: The Harmonizers 2}
            All DW-MRI were processed using the PreQual pipeline to remove eddy current, motion, and EPI distortions \citep{Cai2021}. 
            This submission harmonized across acquisitions by registering the scans of the same subject together, spatial resampling to common domain (acquisition B) and correcting for signal differences with voxel-wise mapping. Average b=0 images of the subject’s diffusion data of acquisitions A and B were mapped to the subject’s T1 image with boundary-based rigid registration \citep{Greve2009} after running FreeSurfer \citep{Fischl2012} on the T1-weighted images. The obtained mappings were used to map the subject’s acquisition A data to the respective acquisition B data, where the mapped data was re-grided to the acquisition B target resolution. The neural network architecture was built to accommodate flexible inputs in a voxel-wise manner: only data from acquisition A, only data from acquisition B, or both. This architecture consisted of an encoder for site A data and a separate encoder for site B data. Both encoders were MLPs with two hidden layers (128 units) with batch normalization and a ReLU activation as well as a linear output layer (95 units). If both A and B data were presented, they were passed through the respective encoder, and the outputs were combined using the mean, yielding a single latent vector. If only A or B data were presented, the data were passed through the respective encoder yielding a latent vector without the need for combination. Regardless of the number of inputs, this latent vector was passed through a decoder with two hidden layers (128 units) with batch normalization and a ReLU activation, as well as one linear output layer (95 units). For every batch, predictions and latent vectors were obtained for a forward pass of A only, B only, and A and B jointly. The closeness of the three latent vectors was incentivized by summing up the pairwise mean squared differences of these latent vectors (latent loss). To incentivize accurate predictions, the mean squared differences of the prediction from A, the prediction from B, and the prediction from A and B to the target data B were summed up (reconstruction loss). Finally, the latent and reconstruction losses were added. The network was trained for 100 epochs with an initial learning rate of 0.1 which was reduced every 25 epochs, i.e. multiplied by 0.1.
            
            \subsubsection{Submission 8: SimpleHarmonics 2}
            All DW-MRI were processed using the PreQual pipeline to remove eddy current, motion, and EPI distortions  \citep{Cai2021}. 
            To harmonize the data, this submission utilized the classic Laplacian approach of spherical harmonic modeling  \citep{Garyfallidis2014} to extract and correct the signal of the 'target'  and ‘reference’ acquisitions. Spherical harmonic representations (order 6) were fit to all DW-MRI. Then, the coefficients (SH coefs) were linearly multiplied with the gradient table of the target site, or acquisition B, to obtain the diffusion signal. The original spacing and number of slices were made consistent with a 3D resample from the AFNI toolkit  \citep{Cox1996}. 
            \subsubsection{Submission 9: SimpleHarmonics 1}
            This submission used the same methods as Submission 1, but the 'target' is Acquisition A, and the 'reference' is Acquisition B. 

    \subsection{Evaluating harmonization}
    Successful harmonization has two main components: reducing acquisition-related bias, and preserving biological variability. We evaluate acquisition related bias by computing Cohen’s D between measures derived from acquisition A and B, as well as the Wilcoxon ranksum test to compare medians (p-value of 0.05 is considered significant). Together these tests inform us if the distributions are significantly different and to what extent. Cohen’s D is interpreted as follows: standardized effect-size of 0.2 is small, 0.5 is medium, and above 0.8 is large \citep{Cohen1977}. 

    We evaluate cross-acquisition bundle shape similarity with the BUAN bundle shape similarity score, a bounded metric between 0 (no similarity) and 1 (near perfect similarity). The BUAN bundle similarity score is a graph-theoretic approach to compare the shapes of two bundles of the same type using bundle adjacency metrics \citep{Chandio2020, Garyfallidis2012}. The BUAN shape similarity score between two bundles is determined by how close two bundles are by computing the minimum flip distance between their respective streamlines and constructing two bundles' coverage of each other. 
    
    For the best-performing harmonization method, we further assess how well it maintains biological variation and the similarity of cross-acquisition bundle shapes. We evaluate biological variation by comparing coefficient of variation (CoV) of the baseline from each acquisition, and CoV after harmonization. Ideally, image acquisition protocol should not influence inter-subject variation. Cross-acquisition differences in CoV suggests there are sources of variation beyond biological that are influencing measurements. Further, effective harmonization would maintain biological variation (CoV should not collapse to near zero), and that between subject variation should be consistent across acquisitions. 

    Competition rankings were based on the average ICC for each metric category (connectomics, microstructure, macrostructure). ChallengeR is an open-source toolkit for analyzing and visualizing challenge results \citep{Wiesenfarth2021}. Using this toolkit, we provide a detailed benchmarking report of each metric individually as supplementary information. 
    
\section{Results}
\subsection{Connectomics}
We evaluate each submission on its ability to harmonize complex network features of the connectome (Figure 3). Submissions 1, 2, and 3 successfully removed significant confounding acquisition effects in all 12 complex network measures and reduced the effect-size difference to “small”. Submissions 5, 8 and 9 were generally unsuccessful in removing this effect, and for density, global and local efficiency, and path length these methods introduced cross acquisition differences that were not in the baseline measures.
\begin{figure}
    \centering
    \includegraphics[width=1.0\linewidth]{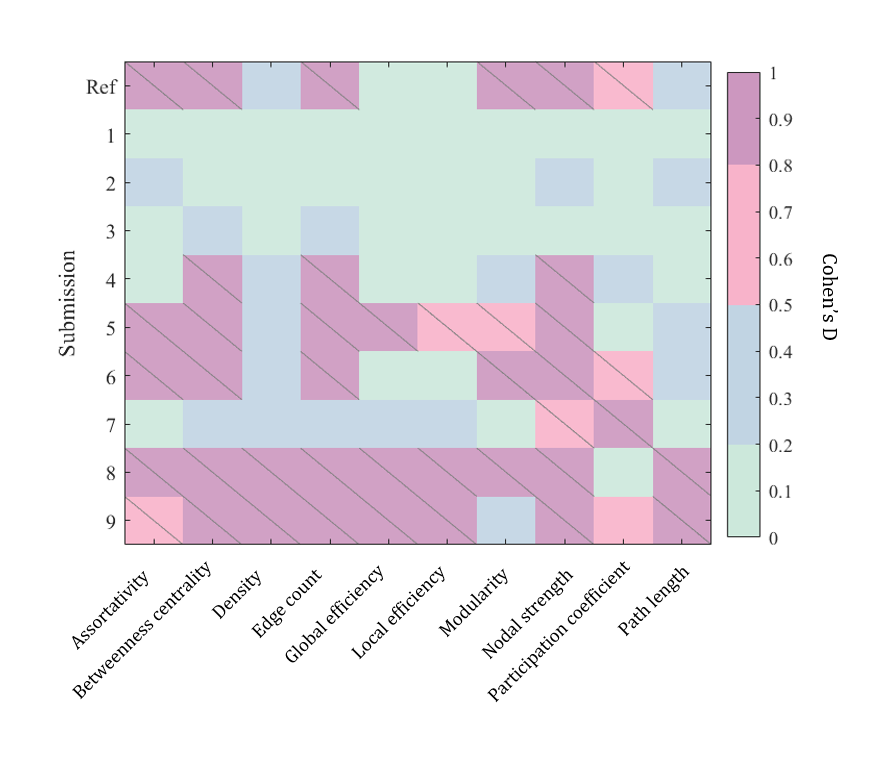}
    \caption{Successful harmonization methods will reduce significant acquisition effects in these measures from the un-harmonized reference (“Ref”). Slashes indicate significant difference (p $\textless$ 0.05) in median between measures derived from acquisitions A and B. We compute Cohen’s D effect-size differences between connectomics measures from acquisitions A and B. }
\end{figure}
\subsection{Bundle macrostructure}
We evaluate the difference in cross-acquisition macrostructure measurements of the same subjects with Cohen's D and the Mann-Whitney U-test (Figure 4). All but submission 1 failed to remove acquisition bias from all measures for bilateral arcuate fasciculus. Across all bundles and macrostructural measures studied, Harmonizers 1 reduced Cohen's D and removed significant median acquisition bias. Submissions 8 and 9 were unsuccessful at harmonizing these measures and introduced significant biases instead. 
We study The Harmonizers 1 bundle macrostructure harmonization more deeply in Figure 5. BUAN bundle similarity score is a graph-theoretic approach to compare the shapes of two bundles of the same type using bundle adjacency metrics \citep{Chandio2020}. We interpret higher values (closer to 1) as extremely close in shape and 0 as no shape similarity. While no reconstructions achieve perfect agreement, after The Harmonizers 1's harmonization is applied, cross-acquisition agreement increases in all six bundles except for AF left. Additionally, we compute BUAN bundle shape similarity scores across the entire test dataset for all submissions (supplementary Figure 1). 
    \begin{figure}[h]
		\centering
		\includegraphics[width=1.0\linewidth]{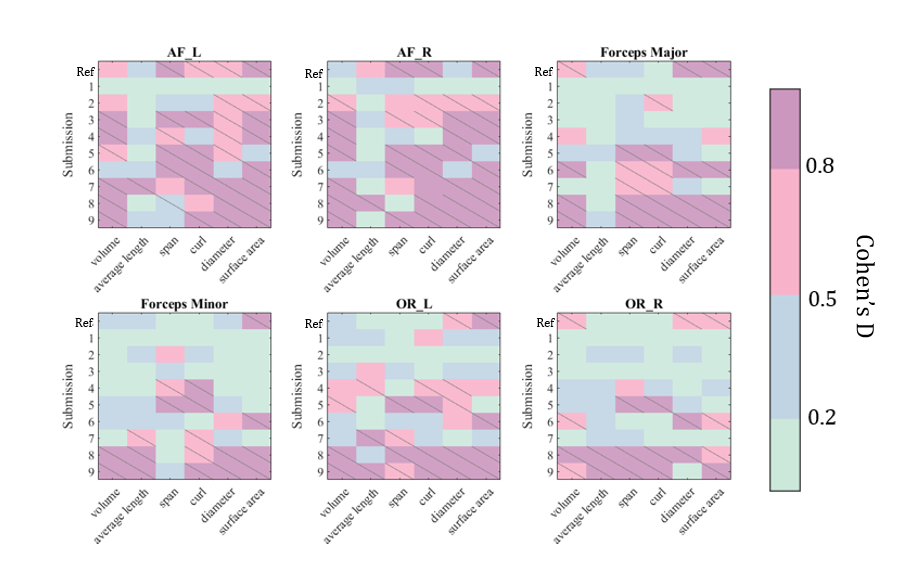}
		\caption{We evaluate each submission on their ability to harmonize macrostructural features of 6 bundles. Successful harmonization will reduce significant acquisition effects in these features from the reference (“Ref”). We report normalized effect-size with Cohen’s D. Slashes indicate a significant difference (p $<$ 0.05) in median between features derived from acquisitions A and B.}
    \end{figure}

    \begin{figure}
    \centering
    \includegraphics[width=0.9\linewidth]{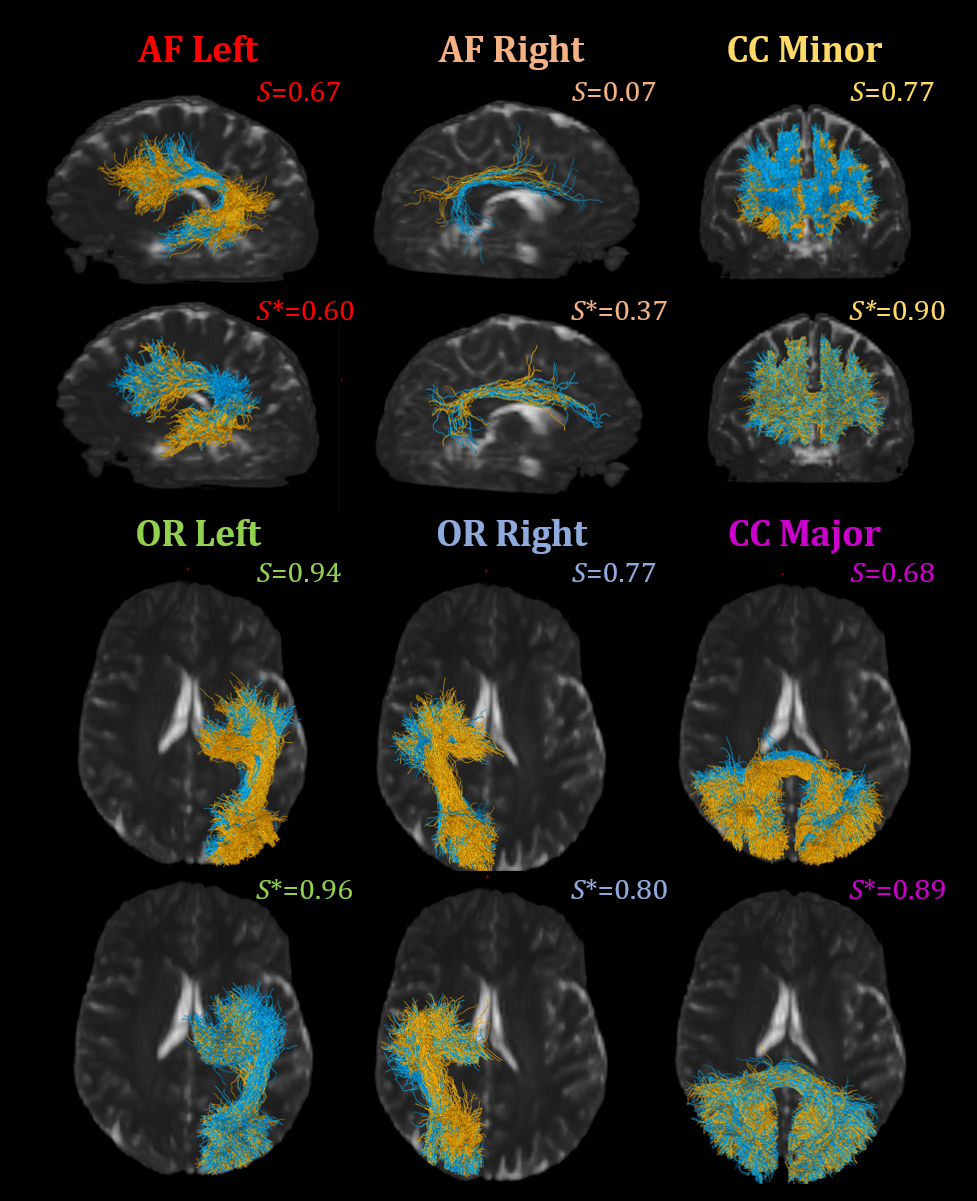}
    \caption{We compare the cross-acquisition (Acquisition A is blue, Acquisition B is orange) shape agreement of reconstructed bundles for one subject in the un-harmonized reference dataset and top performing harmonization technique (The Harmonizers 1) with the BUAN shape similarity metric. Shape similarity scores are reported in the upper right corner of each visualization, with $S$ being the BUAN shape similarity for reference data and $S$* the BUAN shape similarity for harmonized data. Bundle shape similarity score ranges from 0 to 1. The higher the value, the more similar the two bundles are in shape, and the lower value suggests low shape similarity.}
\end{figure}
\subsection{Bundle microstructure}
We evaluate the differences in cross-acquisition microstructure measurements of the same subjects with Cohen's D and the Mann-Whitney U-test (Figure 6). Overall, microstructure had the least number of significant acquisition differences in AD, FA, MD, and RD across the size bundles studied. FA was the only measure that had significant differences in median in half of the bundles (arcuate fascicles right, optical radiation left, and optical radiation right). In fact, all teams except the Harmonizers 1 introduced differences where there were none to ameliorate. Across all bundles and microsructure measures studied, Harmonizers 1 reduced Cohen's D and removed significant median acquisition bias. 

    \begin{figure}[h]
		\centering
		\includegraphics[width=1.0\linewidth]{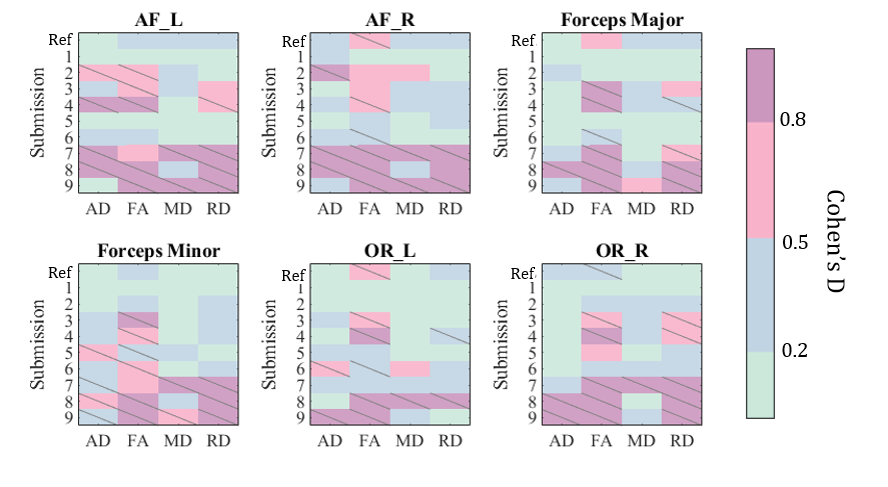}
		\caption{We evaluate each submission on their ability to harmonize microstructural features of 6 bundles. Successful harmonization will reduce significant acquisition effects in these features from the un-harmonized reference (“Ref”). We report normalized effect-size with Cohen’s D. Slashes indicate significant difference (p < 0.05) in median between features derived from acquisitions A and B.}
    \end{figure}
\subsection{Preserving biological differences}
\begin{figure}[h]
		\centering
		\includegraphics[width=0.9\linewidth]{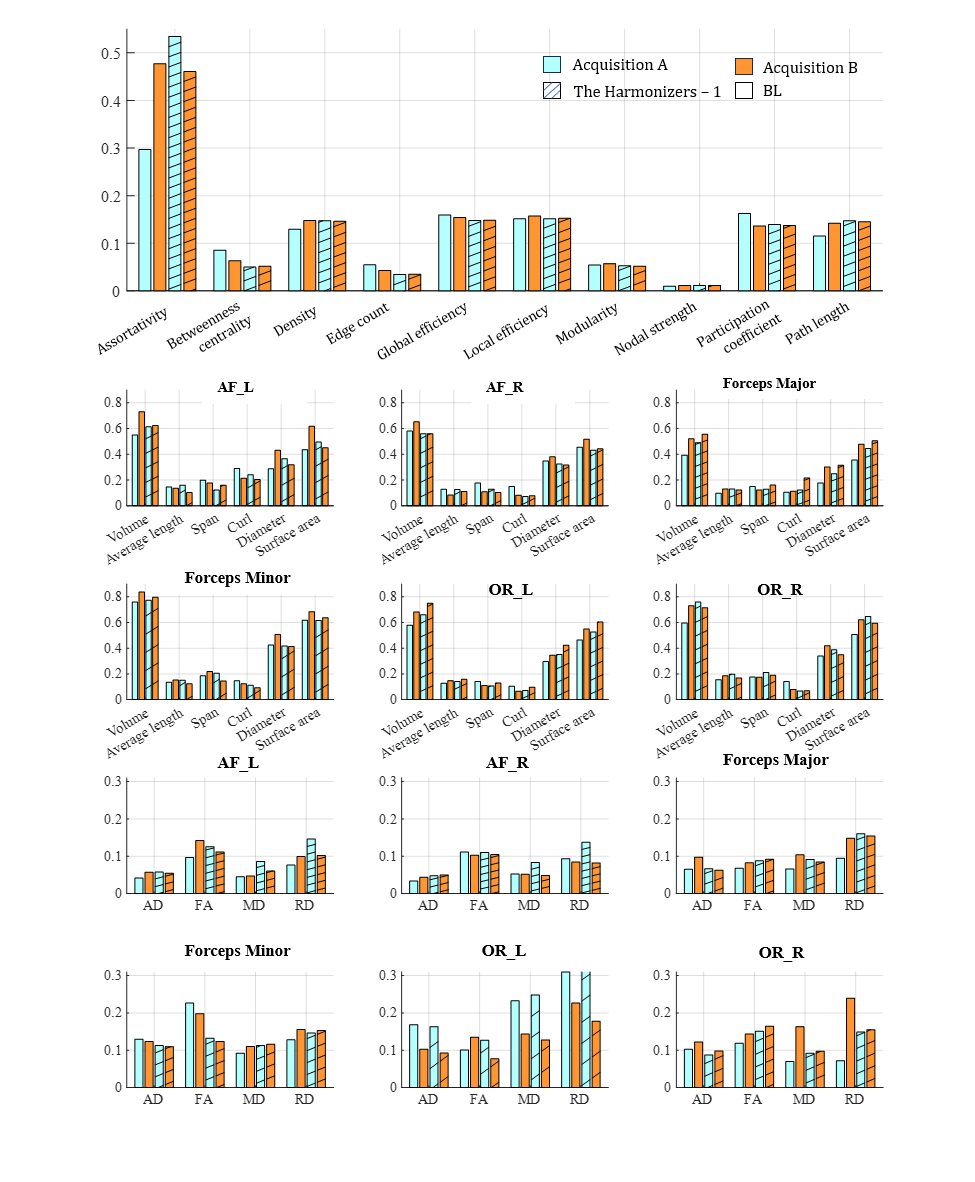}
		\caption{We compute CoV from The Harmonizers 1 and the baseline (BL) connectomics, macrostructure, and microstructure measures for each acquisition.}
\end{figure}

Successful harmonization reduces acquisition bias while preserving biological variance and observed variance should not differ with acquisition protocol. We evaluated the top-performing submission, The Harmonizers 1, based on the amount of variance preserved after harmonization and the agreement between acquisitions (Figure 7). Similar analyses were conducted on NIMG (Supplementary Figure 2) and NeSH (Supplementary Figure 3). 
We found that in the baseline data, inter-subject variation was not equal across acquisitions for any measures studied (Figure 7). Neither acquisition’s CoV was consistently higher or lower than the other across all measures studied. Generally, The Harmonizers 1 (Figure 7), NIMG (Supplementary Figure 2), and NeSH (Supplementary Figure 3) made progress toward equalizing cross-acquisition CoV and maintained inter-subject variation. In none of the methods do we see a dramatic decrease in CoV after harmonization is applied.
We note the following exceptions. Data harmonized using The Harmonizers 1 methodology showed worse agreement of cross-acquisition inter-subject variation in the following measures: average length, span, and RD of AF left bundles; curl in forceps major bundles; and average length, span, diameter, FA, MD, and RD in OR left bundles (Figure 7). Notably, average length, span, curl, FA, MD, AD, and RD exhibited lower overall variation compared to volume, diameter, and surface area, a finding consistent across the six bundles.

\section{Discussion}
In this study we isolate acquisition related differences in a DW-MRI cohort. Specifically, our data includes two distinct angular and spatial resolutions that impact the quality of subsequent models and measurements. To continue dissecting site effects in future studies, a more challenging cohort would contain DW-MRI collected on different scanners and more than two sites.  
Participants of the QuantConn challenge approached the harmonization problem in a variety of ways. The NeSH method targeted and ameliorated the known acquisition differences (spatial and angular resolution). Other methods chose to learn the complex culmination of those differences on the signal (The Harmonizers – 1, DiffusionMaRines, The Harmonizers – 2) or derivative representation (NIMG, NACIS, SimpleHarmonics – 1, SimpleHarmonics – 2). In many real-world cases, acquisition or scanner differences are complex and difficult to disentangle to ameliorate individually. 
future DW-MRI harmonization studies can consider more complex architectures (Cycle-GAN \citep{Hansen2022, Zhu2017}, StyleGAN \citep{Karras2019}) or adaptions of novel MRI harmonization methods \citep{Hu2023, Zuo2021}.

\section{Conclusion}
As a field, we are working toward quantitative analysis of DW-MRI. This necessitates connectomic and tractometry measurements that are robust and reproducible. In this challenge we studied current methods for harmonization of data such that measurements minimize bias from acquisition heterogeneity. As the number of international datasets combining dissimilar acquisitions, scanner manufacturers, and gradient coils grows, so does the need for correction methods to make such data comparable. We find that the most effective harmonization method corrects for motion, eddy-current, and gibbs-ringing distortions, is registered across acquisitions, and voxel-wise signal predicted using an MLP.  The top three performing methods were the Harmonizers – 1, NIMG, and NeSH. We find the following drawbacks and limitations. The Harmonizers – 1 implemented a MLP that requires the same subjects to be scanned with each acquisition protocol and co-registered. Such non-linear registration introduces error (Bierbrier et al., 2022) and warps underlying subject-specific macrostructures. NIMG’s harmonization method does not require perfectly matched pairs, but does rely on non-linear intra- and inter- subject registration of RISH feature maps. This method of mapping RISH features to a common site was also successful in a previous multi-scanner harmonization challenge \citep{Ning2020}. NeSH methodology, notably, does not rely on registration or matched subjects. 


\acks{The Queensland Twin Imaging study data collection was funded by National Institute of Child Health and Human Development (R01 HD050735), and National Health and Medical Research Council (496682, 1009064). This work was also supported in part by NIH grants R01EB017230, NIDDK K01EB032898, R01NS123378, P50HD105353, RF1AG057892 (FiberNET project grant), R01 MH134004 and R01MH119222 (PIs: Rathi, O’Donnell). \\
Tomasz Pieciak acknowledges the Polish National Agency for Academic Exchange for grant PPN/BEK/2019/1/00421 under the Bekker programme and the Ministry of Science and Higher Education (Poland) under the scholarship for outstanding young scientists (692 / STYP /13/2018). \\Tomasz Pieciak, Antonio Tristán Vega and Santiago Aja-Fernández were supported by research grants PID2021-124407NBI00, funded by MCIN/AEI/ 10.13039/ 501100011033 /FEDER, UE, and TED2021-130758B-I00, funded by MCIN /AEI/ 10.13039/ 501100011033 and the European Union “NextGenerationEU/PRTR”. \\
Dominika Ciupek, Maciej Malawski and Julia Machnio were supported by the European Union’s Horizon 2020 research and innovation program under grant agreement Sano No 857533 and the International Research Agendas program of the Foundation for Polish Science No MAB PLUS/2019/13. \\ JYMY and SG received positional funding from the Royal Children’s Hospital Foundation (RCHF 2022-1402). CEK was supported by an Australian Government Research Training Program (RTP) Scholarship, Monash University (Monash Graduate Excellence Scholarship), and the Australian National Health and Medical Research Council (NHMRC; Centre of Research Excellence in Newborn Medicine 1153176). JYMY, SG, JC and CEK acknowledge the support of the Royal Children’s Hospital, Murdoch Children’s Research Institute, The University of Melbourne Department of Paediatrics, and the Victorian Government’s Operational Infrastructure Support Program. Ye Wu is in part supported by the National Key Research and Development Program of China (No. 2023YFF1204800) and the National Natural Science Foundation of China (No. 62201265). Jianzhong He is supported by the National Natural Science Foundation of China (No. 62303413). Fan Zhang is in part supported by the National Key Research and Development Program of China (No. 2023YFE0118600) and the National Natural Science Foundation of China (No. 62371107). The content is solely the responsibility of the authors and does not necessarily represent the official views of the NIH.}

%
\ethics{The work follows appropriate ethical standards in conducting research and writing the manuscript, following all applicable laws and regulations regarding treatment of animals or human subjects.}

\coi{We declare we don't have conflicts of interest.}

\data{Testing and Training data used in this challenge are publicly available at shared Box: https://vanderbilt.app.box.com/s/owijt2mo2vhrp3rjonf90n3hoinygm8z/folder/208448607516}

\bibliography{main}

\begin{thebibliography}{47}
\providecommand{\natexlab}[1]{#1}
\providecommand{\url}[1]{\texttt{#1}}
\expandafter\ifx\csname urlstyle\endcsname\relax
  \providecommand{\doi}[1]{doi: #1}\else
  \providecommand{\doi}{doi: \begingroup \urlstyle{rm}\Url}\fi

\bibitem[NAC()]{NACC}
National alzheimer's coordinating center.
\newblock URL \url{https://naccdata.org/}.

\bibitem[sci()]{scilpy}
scilus/scilpy: The sherbrooke connectivity imaging lab (scil) python dmri processing toolbox.
\newblock URL \url{https://github.com/scilus/scilpy}.

\bibitem[Annet(1970)]{ANNETT1970}
Marion Annet.
\newblock A classification of hand preference by association analysis.
\newblock \emph{British Journal of Psychology}, 61:\penalty0 303--321, 1970.
\newblock ISSN 20448295.
\newblock \doi{10.1111/J.2044-8295.1970.TB01248.X}.

\bibitem[Basser et~al.(1994)Basser, Mattiello, and Lebihan]{Basser1994}
Peter~J. Basser, James Mattiello, and Denis Lebihan.
\newblock Estimation of the effective self-diffusion tensor from the nmr spin echo.
\newblock \emph{Journal of magnetic resonance. Series B}, 103:\penalty0 247--254, 1994.
\newblock ISSN 1064-1866.
\newblock \doi{10.1006/JMRB.1994.1037}.
\newblock URL \url{https://pubmed.ncbi.nlm.nih.gov/8019776/}.

\bibitem[Billah et~al.()Billah, Bouix, Karayumak, and Rathi]{linearrish}
Tashrif Billah, Sylvain Bouix, Suheyla~Cetin Karayumak, and Yogesh Rathi.
\newblock pnlbwh/dmriharmonization: Multi-site dmri harmonization.
\newblock URL \url{https://github.com/pnlbwh/dMRIharmonization}.

\bibitem[Cai et~al.(2021)Cai, Yang, Hansen, Nath, Ramadass, Johnson, Conrad, Boyd, Begnoche, Beason-Held, Shafer, Resnick, Taylor, Price, Morgan, Rogers, Schilling, and Landman]{Cai2021}
Leon~Y. Cai, Qi~Yang, Colin~B. Hansen, Vishwesh Nath, Karthik Ramadass, Graham~W. Johnson, Benjamin~N. Conrad, Brian~D. Boyd, John~P. Begnoche, Lori~L. Beason-Held, Andrea~T. Shafer, Susan~M. Resnick, Warren~D. Taylor, Gavin~R. Price, Victoria~L. Morgan, Baxter~P. Rogers, Kurt~G. Schilling, and Bennett~A. Landman.
\newblock Prequal: An automated pipeline for integrated preprocessing and quality assurance of diffusion weighted mri images.
\newblock \emph{Magnetic resonance in medicine}, 86:\penalty0 456--470, 7 2021.
\newblock ISSN 1522-2594.
\newblock \doi{10.1002/MRM.28678}.
\newblock URL \url{https://pubmed.ncbi.nlm.nih.gov/33533094/}.

\bibitem[Chandio et~al.(2020)Chandio, Risacher, Pestilli, Bullock, Yeh, Koudoro, Rokem, Harezlak, and Garyfallidis]{Chandio2020}
Bramsh~Qamar Chandio, Shannon~Leigh Risacher, Franco Pestilli, Daniel Bullock, Fang-Cheng Yeh, Serge Koudoro, Ariel Rokem, Jaroslaw Harezlak, and Eleftherios Garyfallidis.
\newblock Bundle analytics, a computational framework for investigating the shapes and profiles of brain pathways across populations.
\newblock 2020.
\newblock \doi{10.1038/s41598-020-74054-4}.
\newblock URL \url{https://doi.org/10.1038/s41598-020-74054-4}.

\bibitem[Cohen()]{Cohen1977}
Jacob Cohen.
\newblock Statistical power analysis for the behavioral sciences second edition.

\bibitem[Cox(1996)]{Cox1996}
Robert~W. Cox.
\newblock Afni: Software for analysis and visualization of functional magnetic resonance neuroimages.
\newblock \emph{Computers and Biomedical Research}, 29:\penalty0 162--173, 6 1996.
\newblock ISSN 0010-4809.
\newblock \doi{10.1006/CBMR.1996.0014}.

\bibitem[Dewey et~al.(2019)Dewey, Zhao, Reinhold, Carass, Fitzgerald, Sotirchos, Saidha, Oh, Pham, Calabresi, van Zijl, and Prince]{Dewey2019}
Blake~E. Dewey, Can Zhao, Jacob~C. Reinhold, Aaron Carass, Kathryn~C. Fitzgerald, Elias~S. Sotirchos, Shiv Saidha, Jiwon Oh, Dzung~L. Pham, Peter~A. Calabresi, Peter~C.M. van Zijl, and Jerry~L. Prince.
\newblock Deepharmony: A deep learning approach to contrast harmonization across scanner changes.
\newblock \emph{Magnetic resonance imaging}, 64:\penalty0 160--170, 12 2019.
\newblock ISSN 1873-5894.
\newblock \doi{10.1016/J.MRI.2019.05.041}.
\newblock URL \url{https://pubmed.ncbi.nlm.nih.gov/31301354/}.

\bibitem[Ferrucci(2008)]{Ferrucci2008}
Luigi Ferrucci.
\newblock The baltimore longitudinal study of aging (blsa): A 50-year-long journey and plans for the future.
\newblock \emph{Journals of Gerontology - Series A Biological Sciences and Medical Sciences}, 63:\penalty0 1416--1419, 2008.
\newblock ISSN 10795006.
\newblock \doi{10.1093/GERONA/63.12.1416}.
\newblock URL \url{/record/2009-01339-015}.

\bibitem[Fischl(2012)]{Fischl2012}
Freesurfer~Bruce Fischl.
\newblock Freesurfer.
\newblock 2012.
\newblock \doi{10.1016/j.neuroimage.2012.01.021}.

\bibitem[Garyfallidis et~al.(2012)Garyfallidis, Brett, Correia, Williams, and Nimmo-Smith]{Garyfallidis2012}
Eleftherios Garyfallidis, Matthew Brett, Marta~Morgado Correia, Guy~B. Williams, and Ian Nimmo-Smith.
\newblock Quickbundles, a method for tractography simplification.
\newblock \emph{Frontiers in Neuroscience}, 6, 2012.
\newblock \doi{10.3389/FNINS.2012.00175/FULL}.

\bibitem[Garyfallidis et~al.(2014)Garyfallidis, Brett, Amirbekian, Rokem, van~der Walt, Descoteaux, and Nimmo-Smith]{Garyfallidis2014}
Eleftherios Garyfallidis, Matthew Brett, Bagrat Amirbekian, Ariel Rokem, Stefan van~der Walt, Maxime Descoteaux, and Ian Nimmo-Smith.
\newblock Dipy, a library for the analysis of diffusion mri data.
\newblock \emph{Frontiers in Neuroinformatics}, 8, 2 2014.
\newblock ISSN 16625196.
\newblock \doi{10.3389/FNINF.2014.00008/ABSTRACT}.

\bibitem[Garyfallidis et~al.(2018)Garyfallidis, Côté, Rheault, Sidhu, Hau, Petit, Fortin, Cunanne, and Descoteaux]{Garyfallidis2018}
Eleftherios Garyfallidis, Marc~Alexandre Côté, Francois Rheault, Jasmeen Sidhu, Janice Hau, Laurent Petit, David Fortin, Stephen Cunanne, and Maxime Descoteaux.
\newblock Recognition of white matter bundles using local and global streamline-based registration and clustering.
\newblock \emph{NeuroImage}, 170:\penalty0 283--295, 4 2018.
\newblock ISSN 1095-9572.
\newblock \doi{10.1016/J.NEUROIMAGE.2017.07.015}.
\newblock URL \url{https://pubmed.ncbi.nlm.nih.gov/28712994/}.

\bibitem[Greve and Fischl(2009)]{Greve2009}
Douglas~N. Greve and Bruce Fischl.
\newblock Accurate and robust brain image alignment using boundary-based registration.
\newblock \emph{NeuroImage}, 48:\penalty0 63--72, 10 2009.
\newblock ISSN 1095-9572.
\newblock \doi{10.1016/J.NEUROIMAGE.2009.06.060}.
\newblock URL \url{https://pubmed.ncbi.nlm.nih.gov/19573611/}.

\bibitem[Hansen et~al.(2022)Hansen, Schilling, Rheault, Resnick, Shafer, Beason-Held, and Landman]{Hansen2022}
Colin~B. Hansen, Kurt~G. Schilling, Francois Rheault, Susan Resnick, Andrea~T. Shafer, Lori~L. Beason-Held, and Bennett~A. Landman.
\newblock Contrastive semi-supervised harmonization of single-shell to multi-shell diffusion mri.
\newblock \emph{Magnetic Resonance Imaging}, 93:\penalty0 73--86, 11 2022.
\newblock ISSN 0730-725X.
\newblock \doi{10.1016/J.MRI.2022.06.004}.

\bibitem[Hendriks et~al.(2023)Hendriks, Vilanova, and Chamberland]{HendricksCDMRI2023}
Tom Hendriks, Anna Vilanova, and Maxime Chamberland.
\newblock Neural spherical harmonics for structurally coherent continuous representation of diffusion mri signal.
\newblock In Muge Karaman, Remika Mito, Elizabeth Powell, Francois Rheault, and Stefan Winzeck, editors, \emph{Computational Diffusion MRI}, pages 1--12, Cham, 2023. Springer Nature Switzerland.
\newblock ISBN 978-3-031-47292-3.

\bibitem[Hu et~al.(2023)Hu, Chen, Horng, Bashyam, Davatzikos, Alexander-Bloch, Li, Shou, Satterthwaite, Yu, and Shinohara]{Hu2023}
Fengling Hu, Andrew~A. Chen, Hannah Horng, Vishnu Bashyam, Christos Davatzikos, Aaron Alexander-Bloch, Mingyao Li, Haochang Shou, Theodore~D. Satterthwaite, Meichen Yu, and Russell~T. Shinohara.
\newblock Image harmonization: A review of statistical and deep learning methods for removing batch effects and evaluation metrics for effective harmonization.
\newblock \emph{NeuroImage}, 274:\penalty0 120125, 7 2023.
\newblock ISSN 1053-8119.
\newblock \doi{10.1016/J.NEUROIMAGE.2023.120125}.

\bibitem[Jack et~al.(2008)Jack, Bernstein, Fox, Thompson, Alexander, Harvey, Borowski, Britson, Whitwell, Ward, Dale, Felmlee, Gunter, Hill, Killiany, Schuff, Fox-Bosetti, Lin, Studholme, DeCarli, Krueger, Ward, Metzger, Scott, Mallozzi, Blezek, Levy, Debbins, Fleisher, Albert, Green, Bartzokis, Glover, Mugler, and Weiner]{Jack2008}
Clifford~R. Jack, Matt~A. Bernstein, Nick~C. Fox, Paul Thompson, Gene Alexander, Danielle Harvey, Bret Borowski, Paula~J. Britson, Jennifer~L. Whitwell, Chadwick Ward, Anders~M. Dale, Joel~P. Felmlee, Jeffrey~L. Gunter, Derek~L.G. Hill, Ron Killiany, Norbert Schuff, Sabrina Fox-Bosetti, Chen Lin, Colin Studholme, Charles~S. DeCarli, Gunnar Krueger, Heidi~A. Ward, Gregory~J. Metzger, Katherine~T. Scott, Richard Mallozzi, Daniel Blezek, Joshua Levy, Josef~P. Debbins, Adam~S. Fleisher, Marilyn Albert, Robert Green, George Bartzokis, Gary Glover, John Mugler, and Michael~W. Weiner.
\newblock The alzheimer's disease neuroimaging initiative (adni): Mri methods.
\newblock \emph{Journal of magnetic resonance imaging : JMRI}, 27:\penalty0 685--691, 4 2008.
\newblock ISSN 1053-1807.
\newblock \doi{10.1002/JMRI.21049}.
\newblock URL \url{https://pubmed.ncbi.nlm.nih.gov/18302232/}.

\bibitem[Jenkinson et~al.(2012)Jenkinson, Beckmann, Behrens, Woolrich, and Smith]{fsl}
Mark Jenkinson, Christian~F. Beckmann, Timothy~E.J. Behrens, Mark~W. Woolrich, and Stephen~M. Smith.
\newblock Fsl.
\newblock \emph{NeuroImage}, 62:\penalty0 782--790, 8 2012.
\newblock ISSN 1095-9572.
\newblock \doi{10.1016/J.NEUROIMAGE.2011.09.015}.
\newblock URL \url{https://pubmed.ncbi.nlm.nih.gov/21979382/}.

\bibitem[Karayumak et~al.(2019)Karayumak, Bouix, Ning, James, Crow, Shenton, Kubicki, and Rathi]{Karayumak2019}
Suheyla~Cetin Karayumak, Sylvain Bouix, Lipeng Ning, Anthony James, Tim Crow, Martha Shenton, Marek Kubicki, and Yogesh Rathi.
\newblock Retrospective harmonization of multi-site diffusion mri data acquired with different acquisition parameters.
\newblock \emph{NeuroImage}, 184:\penalty0 180--200, 1 2019.
\newblock ISSN 10959572.
\newblock \doi{10.1016/j.neuroimage.2018.08.073}.

\bibitem[Karras et~al.(2019)Karras, Laine, and Aila]{Karras2019}
Tero Karras, Samuli Laine, and Timo Aila.
\newblock A style-based generator architecture for generative adversarial networks, 2019.
\newblock URL \url{https://github.com/NVlabs/stylegan}.

\bibitem[Klein and Tourville(2012)]{Klein2012}
Arno Klein and Jason Tourville.
\newblock 101 labeled brain images and a consistent human cortical labeling protocol.
\newblock \emph{Frontiers in Neuroscience}, 0:\penalty0 171, 2012.
\newblock ISSN 16624548.
\newblock \doi{10.3389/FNINS.2012.00171/ABSTRACT}.

\bibitem[LaMontagne et~al.(2019)LaMontagne, Benzinger, Morris, Keefe, Hornbeck, Xiong, Grant, Hassenstab, Moulder, Vlassenko, Raichle, Cruchaga, and Marcus]{LaMontagne2019}
Pamela~J. LaMontagne, Tammie~LS. Benzinger, John~C. Morris, Sarah Keefe, Russ Hornbeck, Chengjie Xiong, Elizabeth Grant, Jason Hassenstab, Krista Moulder, Andrei~G. Vlassenko, Marcus~E. Raichle, Carlos Cruchaga, and Daniel Marcus.
\newblock Oasis-3: Longitudinal neuroimaging, clinical, and cognitive dataset for normal aging and alzheimer disease.
\newblock \emph{medRxiv}, page 2019.12.13.19014902, 12 2019.
\newblock \doi{10.1101/2019.12.13.19014902}.
\newblock URL \url{https://www.medrxiv.org/content/10.1101/2019.12.13.19014902v1 https://www.medrxiv.org/content/10.1101/2019.12.13.19014902v1.abstract}.

\bibitem[Mirzaalian et~al.(2018)Mirzaalian, Ning, Savadjiev, Pasternak, Bouix, Michailovich, Karmacharya, Grant, Marx, Morey, Flashman, George, McAllister, Andaluz, Shutter, Coimbra, Zafonte, Coleman, Kubicki, Westin, Stein, Shenton, and Rathi]{Mirzaalian2018}
Hengameh Mirzaalian, Lipeng Ning, Peter Savadjiev, Ofer Pasternak, Sylvain Bouix, Oleg Michailovich, Sarina Karmacharya, Gerald Grant, Christine~E. Marx, Rajendra~A. Morey, Laura~A. Flashman, Mark~S. George, Thomas~W. McAllister, Norberto Andaluz, Lori Shutter, Raul Coimbra, Ross~D. Zafonte, Mike~J. Coleman, Marek Kubicki, Carl~Fredrik Westin, Murray~B. Stein, Martha~E. Shenton, and Yogesh Rathi.
\newblock Multi-site harmonization of diffusion mri data in a registration framework.
\newblock \emph{Brain Imaging and Behavior}, 12:\penalty0 284--295, 2 2018.
\newblock ISSN 19317565.
\newblock \doi{10.1007/S11682-016-9670-Y/TABLES/8}.
\newblock URL \url{https://link.springer.com/article/10.1007/s11682-016-9670-y}.

\bibitem[Nencka et~al.(2018)Nencka, Meier, Wang, Muftuler, Wu, Saykin, Harezlak, Brooks, Giza, Difiori, Guskiewicz, Mihalik, LaConte, Duma, Broglio, McAllister, McCrea, and Koch]{Nencka2018}
Andrew~S. Nencka, Timothy~B. Meier, Yang Wang, L.~Tugan Muftuler, Yu~Chien Wu, Andrew~J. Saykin, Jaroslaw Harezlak, M.~Alison Brooks, Christopher~C. Giza, John Difiori, Kevin~M. Guskiewicz, Jason~P. Mihalik, Stephen~M. LaConte, Stefan~M. Duma, Steven Broglio, Thomas McAllister, Michael~A. McCrea, and Kevin~M. Koch.
\newblock Stability of mri metrics in the advanced research core of the ncaa-dod concussion assessment, research and education (care) consortium.
\newblock \emph{Brain imaging and behavior}, 12:\penalty0 1121--1140, 8 2018.
\newblock ISSN 1931-7565.
\newblock \doi{10.1007/S11682-017-9775-Y}.
\newblock URL \url{https://pubmed.ncbi.nlm.nih.gov/29064019/}.

\bibitem[Newlin et~al.(2023)Newlin, Rheault, Schilling, and Landman]{Newlin2023}
Nancy~R. Newlin, François Rheault, Kurt~G. Schilling, and Bennett~A. Landman.
\newblock Characterizing streamline count invariant graph measures of structural connectomes.
\newblock \emph{Journal of Magnetic Resonance Imaging}, 2023.
\newblock ISSN 1522-2586.
\newblock \doi{10.1002/JMRI.28631}.
\newblock URL \url{https://onlinelibrary.wiley.com/doi/full/10.1002/jmri.28631 https://onlinelibrary.wiley.com/doi/abs/10.1002/jmri.28631 https://onlinelibrary.wiley.com/doi/10.1002/jmri.28631}.

\bibitem[Ni et~al.(2006)Ni, Kavcic, Zhu, Ekholm, and Zhong]{Ni2006}
H.~Ni, V.~Kavcic, T.~Zhu, S.~Ekholm, and Jianhui Zhong.
\newblock Effects of number of diffusion gradient directions on derived diffusion tensor imaging indices in human brain.
\newblock \emph{AJNR: American Journal of Neuroradiology}, 27:\penalty0 1776, 9 2006.
\newblock ISSN 01956108.
\newblock URL \url{/pmc/articles/PMC8139764/ /pmc/articles/PMC8139764/?report=abstract https://www.ncbi.nlm.nih.gov/pmc/articles/PMC8139764/}.

\bibitem[Ning et~al.(2019)Ning, Bonet-Carne, Grussu, Sepehrband, Kaden, Veraart, Blumberg, Khoo, Palombo, Coll-Font, Scherrer, Warfield, Karayumak, Rathi, Koppers, Weninger, Ebert, Merhof, Moyer, Pietsch, Christiaens, Teixeira, Tournier, Zhylka, Pluim, Parker, Rudrapatna, Evans, Charron, Jones, and Tax]{Ning2019}
Lipeng Ning, Elisenda Bonet-Carne, Francesco Grussu, Farshid Sepehrband, Enrico Kaden, Jelle Veraart, Stefano~B. Blumberg, Can~Son Khoo, Marco Palombo, Jaume Coll-Font, Benoit Scherrer, Simon~K. Warfield, Suheyla~Cetin Karayumak, Yogesh Rathi, Simon Koppers, Leon Weninger, Julia Ebert, Dorit Merhof, Daniel Moyer, Maximilian Pietsch, Daan Christiaens, Rui Teixeira, Jacques~Donald Tournier, Andrey Zhylka, Josien Pluim, Greg Parker, Umesh Rudrapatna, John Evans, Cyril Charron, Derek~K. Jones, and Chantal~W.M. Tax.
\newblock Muti-shell diffusion mri harmonisation and enhancement challenge (mushac): Progress and results.
\newblock \emph{Mathematics and Visualization}, pages 217--224, 2019.
\newblock ISSN 2197666X.
\newblock \doi{10.1007/978-3-030-05831-9_18/COVER}.
\newblock URL \url{https://link.springer.com/chapter/10.1007/978-3-030-05831-9_18}.

\bibitem[Ning et~al.(2020)Ning, Bonet-Carne, Grussu, Sepehrband, Kaden, Veraart, Blumberg, Khoo, Palombo, Kokkinos, Alexander, Coll-Font, Scherrer, Warfield, Karayumak, Rathi, Koppers, Weninger, Ebert, Merhof, Moyer, Pietsch, Christiaens, Teixeira, Tournier, Schilling, Huo, Nath, Hansen, Blaber, Landman, Zhylka, Pluim, Parker, Rudrapatna, Evans, Charron, Jones, and Tax]{Ning2020}
Lipeng Ning, Elisenda Bonet-Carne, Francesco Grussu, Farshid Sepehrband, Enrico Kaden, Jelle Veraart, Stefano~B. Blumberg, Can~Son Khoo, Marco Palombo, Iasonas Kokkinos, Daniel~C. Alexander, Jaume Coll-Font, Benoit Scherrer, Simon~K. Warfield, Suheyla~Cetin Karayumak, Yogesh Rathi, Simon Koppers, Leon Weninger, Julia Ebert, Dorit Merhof, Daniel Moyer, Maximilian Pietsch, Daan Christiaens, Rui Azeredo~Gomes Teixeira, Jacques~Donald Tournier, Kurt~G. Schilling, Yuankai Huo, Vishwesh Nath, Colin Hansen, Justin Blaber, Bennett~A. Landman, Andrey Zhylka, Josien~P.W. Pluim, Greg Parker, Umesh Rudrapatna, John Evans, Cyril Charron, Derek~K. Jones, and Chantal~M.W. Tax.
\newblock Cross-scanner and cross-protocol multi-shell diffusion mri data harmonization: Algorithms and results.
\newblock \emph{NeuroImage}, 221:\penalty0 117128, 11 2020.
\newblock ISSN 1053-8119.
\newblock \doi{10.1016/J.NEUROIMAGE.2020.117128}.

\bibitem[Onicas et~al.(2022)Onicas, Ware, Harris, Beauchamp, Beaulieu, Craig, Doan, Freedman, Goodyear, Zemek, Yeates, and Lebel]{Onicas2022}
Adrian~I. Onicas, Ashley~L. Ware, Ashley~D. Harris, Miriam~H. Beauchamp, Christian Beaulieu, William Craig, Quynh Doan, Stephen~B. Freedman, Bradley~G. Goodyear, Roger Zemek, Keith~Owen Yeates, and Catherine Lebel.
\newblock Multisite harmonization of structural dti networks in children: An a-cap study.
\newblock \emph{Frontiers in Neurology}, 13, 6 2022.
\newblock \doi{10.3389/FNEUR.2022.850642}.

\bibitem[Pierpaoli et~al.(1996)Pierpaoli, Jezzard, Basser, Barnett, and Chiro]{Pierpaoli1996}
Carlo Pierpaoli, Peter Jezzard, Peter~J. Basser, Alan Barnett, and Giovanni~Di Chiro.
\newblock Diffusion tensor mr imaging of the human brain.
\newblock \emph{https://doi.org/10.1148/radiology.201.3.8939209}, 201:\penalty0 637--648, 12 1996.
\newblock ISSN 00338419.
\newblock \doi{10.1148/RADIOLOGY.201.3.8939209}.
\newblock URL \url{https://pubs.rsna.org/doi/10.1148/radiology.201.3.8939209}.

\bibitem[Pinto et~al.(2020)Pinto, Paolella, Billiet, Dyck, Guns, Jeurissen, Ribbens, den Dekker, and Sijbers]{Pinto2020}
Maíra~Siqueira Pinto, Roberto Paolella, Thibo Billiet, Pieter~Van Dyck, Pieter~Jan Guns, Ben Jeurissen, Annemie Ribbens, Arnold~J. den Dekker, and Jan Sijbers.
\newblock Harmonization of brain diffusion mri: Concepts and methods.
\newblock \emph{Frontiers in neuroscience}, 14, 5 2020.
\newblock ISSN 1662-4548.
\newblock \doi{10.3389/FNINS.2020.00396}.
\newblock URL \url{https://pubmed.ncbi.nlm.nih.gov/32435181/}.

\bibitem[Pizzolato et~al.(2020)Pizzolato, Palombo, Hutter, Nash, Zhang, and Gyori]{Pizzolato2020}
Marco Pizzolato, Marco Palombo, Jana Hutter, Vishwesh Nash, Fan Zhang, and Noemi Gyori.
\newblock Super-resolution of multi dimensional diffusion mri data.
\newblock 3 2020.
\newblock \doi{10.5281/ZENODO.3718990}.
\newblock URL \url{https://zenodo.org/record/3718990}.

\bibitem[Rubinov and Sporns(2010)]{bct}
Mikail Rubinov and Olaf Sporns.
\newblock Complex network measures of brain connectivity: uses and interpretations.
\newblock \emph{NeuroImage}, 52:\penalty0 1059--1069, 9 2010.
\newblock ISSN 1095-9572.
\newblock \doi{10.1016/J.NEUROIMAGE.2009.10.003}.
\newblock URL \url{https://pubmed.ncbi.nlm.nih.gov/19819337/}.

\bibitem[Schilling et~al.(2021)Schilling, Tax, Rheault, Hansen, Yang, Yeh, Cai, Anderson, and Landman]{Schilling2021}
Kurt~G. Schilling, Chantal~M.W. Tax, Francois Rheault, Colin Hansen, Qi~Yang, Fang~Cheng Yeh, Leon Cai, Adam~W. Anderson, and Bennett~A. Landman.
\newblock Fiber tractography bundle segmentation depends on scanner effects, vendor effects, acquisition resolution, diffusion sampling scheme, diffusion sensitization, and bundle segmentation workflow.
\newblock \emph{NeuroImage}, 242:\penalty0 118451, 11 2021.
\newblock ISSN 1053-8119.
\newblock \doi{10.1016/J.NEUROIMAGE.2021.118451}.

\bibitem[Stejskal and Tanner(1965)]{Stejskal1965}
E~O Stejskal and ;~J~E Tanner.
\newblock Spin diffusion measurements: Spin echoes in the presence of a time-dependent field gradient.
\newblock \emph{J. Chem. Phys}, 42:\penalty0 288--292, 1965.
\newblock \doi{10.1063/1.1695690}.
\newblock URL \url{https://doi.org/10.1063/1.1695690}.

\bibitem[Strike et~al.(2023)Strike, Blokland, Hansell, Martin, Toga, Thompson, de~Zubicaray, McMahon, and Wright]{Queensland}
Lachlan~T. Strike, Gabriella~A.M. Blokland, Narelle~K. Hansell, Nicholas~G. Martin, Arthur~W. Toga, Paul~M. Thompson, Greig~I. de~Zubicaray, Katie~L. McMahon, and Margaret~J. Wright, 2023.

\bibitem[Tournier et~al.(2007)Tournier, Calamante, and Connelly]{Tournier2007}
J.~Donald Tournier, Fernando Calamante, and Alan Connelly.
\newblock Robust determination of the fibre orientation distribution in diffusion mri: Non-negativity constrained super-resolved spherical deconvolution.
\newblock \emph{NeuroImage}, 35:\penalty0 1459--1472, 5 2007.
\newblock ISSN 1053-8119.
\newblock \doi{10.1016/J.NEUROIMAGE.2007.02.016}.

\bibitem[Tournier et~al.(2013)Tournier, Calamante, and Connelly]{Tournier2013}
J.~Donald Tournier, Fernando Calamante, and Alan Connelly.
\newblock Determination of the appropriate b value and number of gradient directions for high-angular-resolution diffusion-weighted imaging.
\newblock \emph{NMR in biomedicine}, 26:\penalty0 1775--1786, 12 2013.
\newblock ISSN 1099-1492.
\newblock \doi{10.1002/NBM.3017}.
\newblock URL \url{https://pubmed.ncbi.nlm.nih.gov/24038308/}.

\bibitem[Tournier et~al.(2019)Tournier, Smith, Raffelt, Tabbara, Dhollander, Pietsch, Christiaens, Jeurissen, Yeh, and Connelly]{mrtrix}
J.~Donald Tournier, Robert Smith, David Raffelt, Rami Tabbara, Thijs Dhollander, Maximilian Pietsch, Daan Christiaens, Ben Jeurissen, Chun~Hung Yeh, and Alan Connelly.
\newblock Mrtrix3: A fast, flexible and open software framework for medical image processing and visualisation.
\newblock \emph{NeuroImage}, 202:\penalty0 116137, 11 2019.
\newblock ISSN 1053-8119.
\newblock \doi{10.1016/J.NEUROIMAGE.2019.116137}.

\bibitem[Vollmar et~al.(2010)Vollmar, O'Muircheartaigh, Barker, Symms, Thompson, Kumari, Duncan, Richardson, and Koepp]{Vollmar2010}
Christian Vollmar, Jonathan O'Muircheartaigh, Gareth~J. Barker, Mark~R. Symms, Pamela Thompson, Veena Kumari, John~S. Duncan, Mark~P. Richardson, and Matthias~J. Koepp.
\newblock Identical, but not the same: intra-site and inter-site reproducibility of fractional anisotropy measures on two 3.0t scanners.
\newblock \emph{NeuroImage}, 51:\penalty0 1384--1394, 7 2010.
\newblock ISSN 1095-9572.
\newblock \doi{10.1016/J.NEUROIMAGE.2010.03.046}.
\newblock URL \url{https://pubmed.ncbi.nlm.nih.gov/20338248/}.

\bibitem[Westin et~al.(2002)Westin, Maier, Mamata, Nabavi, Jolesz, and Kikinis]{Westin2002}
C.~F. Westin, S.~E. Maier, H.~Mamata, A.~Nabavi, F.~A. Jolesz, and R.~Kikinis.
\newblock Processing and visualization for diffusion tensor mri.
\newblock \emph{Medical Image Analysis}, 6:\penalty0 93--108, 6 2002.
\newblock ISSN 1361-8415.
\newblock \doi{10.1016/S1361-8415(02)00053-1}.

\bibitem[Wiesenfarth et~al.(2021)Wiesenfarth, Reinke, Landman, Eisenmann, Saiz, Cardoso, Maier-Hein, and Kopp-Schneider]{Wiesenfarth2021}
Manuel Wiesenfarth, Annika Reinke, Bennett~A. Landman, Matthias Eisenmann, Laura~Aguilera Saiz, M.~Jorge Cardoso, Lena Maier-Hein, and Annette Kopp-Schneider.
\newblock Methods and open-source toolkit for analyzing and visualizing challenge results.
\newblock \emph{Scientific Reports 2021 11:1}, 11:\penalty0 1--15, 1 2021.
\newblock ISSN 2045-2322.
\newblock \doi{10.1038/s41598-021-82017-6}.
\newblock URL \url{https://www.nature.com/articles/s41598-021-82017-6}.

\bibitem[Zhu et~al.(2017)Zhu, Park, Isola, and Efros]{Zhu2017}
Jun~Yan Zhu, Taesung Park, Phillip Isola, and Alexei~A. Efros.
\newblock Unpaired image-to-image translation using cycle-consistent adversarial networks.
\newblock \emph{Proceedings of the IEEE International Conference on Computer Vision}, 2017-October:\penalty0 2242--2251, 12 2017.
\newblock ISSN 15505499.
\newblock \doi{10.1109/ICCV.2017.244}.

\bibitem[Zuo et~al.(2021)Zuo, Dewey, Liu, He, Newsome, Mowry, Resnick, Prince, and Carass]{Zuo2021}
Lianrui Zuo, Blake~E. Dewey, Yihao Liu, Yufan He, Scott~D. Newsome, Ellen~M. Mowry, Susan~M. Resnick, Jerry~L. Prince, and Aaron Carass.
\newblock Unsupervised mr harmonization by learning disentangled representations using information bottleneck theory.
\newblock \emph{NeuroImage}, 243:\penalty0 118569, 11 2021.
\newblock ISSN 1053-8119.
\newblock \doi{10.1016/J.NEUROIMAGE.2021.118569}.

\end{thebibliography}


\end{document}